\documentclass[twocolumn,showpacs,preprintnumbers,amsmath,amssymb]{revtex4}

\usepackage{graphicx}
\usepackage{dcolumn}
\usepackage{bm}
\newcommand{\sech}{\mathrm{sech}}
\newcommand{\Ri}{\mathrm{Ri}}

\newcommand{\Ima}{\mathrm{Im}}

\begin{document}


\title{On Holmboe's instability for smooth shear and density profiles}

\author{Alexandros Alexakis}
\affiliation{National Center for Atmospheric Research, Boulder, Colorado, 80305 USA}
\altaffiliation{The National Center for Atmospheric Research is supported by the National 
Science Foundation}

\date{\today}

\begin{abstract}
The linear stability of a stratified shear flow for smooth density
profiles is studied. This work focuses on the nature of the
stability boundaries of flows in which both Kelvin-Helmholtz and
Holmboe instabilities are present. For
a fixed Richardson number the unstable modes are confined
to finite bands between a smallest and a largest marginally unstable
wavenumber. The results in this paper indicate that the stability boundary 
for small wavenumbers is comprised of neutral modes with phase velocity equal to the
maximum/minimum wind velocity  whereas the other stability boundary, for large
wavenumbers, is comprised of singular neutral modes with phase
velocity in the range of the velocity shear. We show how these
stability boundaries can be evaluated without solving for the
growth rate over the entire parameter space as was previously
done. The results indicate further that there is a new instability domain
that has not been previously noted in the literature. The unstable
modes, in this new instability domain, appear for larger values of the Richardson
number and are related to the higher harmonics of the internal
gravity wave spectrum.

\end{abstract}


\maketitle

\section{Introduction}
\label{Intro}

Stratified shear flow instabilities occur 
in a variety of physical contexts such as astrophysics
\cite{Rosner01}, the Earth's atmosphere and oceanography
\cite{Armi88,Oguz90,Sargent87,Yoshida98}. 
The linear instability problem of stratified shear flows
has been addressed in a large body of literature.
Since the original work of Helmholtz \cite{Helmholtz68} and Lord Kelvin
\cite{Kelvin10} many models of flows and density stratifications have
been investigated both analytically and numerically.
Parameterized by the global
Richardson number, these investigations have resulted in a large variety
of stability/instability domains in the Richardson number - wavenumber 
space \cite{Howard63,Howard66,Howard73,Drazin81}.
However, a full understanding of this large variety
of stability domains does not exist.

A step towards understanding a particular aspect of shear flow instability
was achieved by Holmboe \cite{Holmboe62}.
Using a piece-wise-linear form of the velocity
and density profile Holmboe managed to distinguish between two classes
of unstable modes that are present in stratified shear flows.
In the first class the unstable modes have zero phase velocity in
a reference frame of zero-mean velocity and exist for a finite
range of strength of the stratification (Richardson number). 
This instability is
referred to in the literature as Kelvin-Helmholtz instability
since the behavior of the unstable modes resembles the one
predicted by Kelvin-Helmholtz. The unstable modes of the second
class have non-zero phase velocity and typically smaller growth
rate than the Kelvin-Helmholtz modes. However they are present for
arbitrarily large values of strength of the stratification, making
them better candidates to explain certain physical phenomena.
This second kind of instability is referred to as Holmboe
instability.

Several authors have expanded Holmboe's work \cite{Caulfield94,Haigh99,Lawrence91} 
by considering
different stratification and velocity profiles that do not include
the simplifying symmetries Holmboe used in his model. 
Discussions on the mechanisms involved in the instability can
be found in \cite{Caulfield94,Baines94}.
Hazel \cite{Hazel72} and more recently 
Smyth and Peltier \cite{Smyth89} have shown that Holmboe's results hold
for smooth density and velocity profiles as long as the length
scale of the density variation is sufficiently smaller than the
length scale of the velocity variation. 
Further, effects of viscosity and diffusivity 
\citep{Nishida87,Smyth90}, 
non-linear evolution 
\cite{Smyth88,Smyth91,Sutherland94} and
mixing properties \cite{Smyth03} of the Holmboe instability have also been
investigated. 
Experimentally Holmboe's
instability has been investigated by various groups. Browand and Winant
\cite{Browand73}
first performed shear flow experiments in stratified environments
under the conditions that Holmboe's instabilities are present.
Their investigation has been extended further
by more recent experiments 
\citep{Koop76,Lawrence91,Pouliquen01,Zhu01,Hogg03,Yonemitsu96}. 
The unstable
Holmboe modes have been observed and Holmboe's predictions verified. 

In this work we examine the stability of
smoothly stratified shear flows as was done in \cite{Hazel72} and
\cite{Smyth89}, but here we focus
on finding the kind of marginally unstable modes that comprise the
stability boundaries. We show how these modes can be determined without
solving the full eigenvalue problem for the complex eigenvalue
$c$. Furthermore, our results indicate that new instability
regions exist which have not been discovered before in the literature.

This paper is structured as follows.
In section \ref{Form} we formulate the linear stability problem and
discuss the possible marginally unstable modes that can
comprise the stability boundaries.
In section \ref{NumMeth} we describe the numerical methods used.
Section \ref{Res} presents our results. 
Specifically, we show the location of
the marginally unstable modes and numerically verify that they
indeed constitute the stability boundaries.
Conclusions are drawn in section \ref{Cons} where we also give
a physical description of our results.

\section{Formulation}
\label{Form}

We begin with the Taylor-Goldstein equation, which describes
linear normal modes of a parallel shear flow in a stratified,
inviscid, non-diffusive, Boussinesq fluid: 
\begin{equation}
\label{TG}
\phi''
- \left[ k^2 + \frac{U''}{U-c} -\frac{J(y)}{(U-c)^2} \right]\phi=0,
\end{equation}
where $\phi(y)$ is the complex amplitude of the stream function
for a normal mode with real wavenumber $k$ and complex phase velocity $c$.
$\Ima\{c\}>0$ implies instability with growth rate given
by $\gamma=k\Ima\{c\}$. $U(y)$ is the unperturbed velocity in the
$x$ direction and $J(y)=-g\rho^{-1}d\rho/dy$ is the squared
Brunt-V\"ais\"ala frequency. Prime indicates differentiation with
respect to $y$.
We note that when $c$ is real and in the range of $U$
there is a height $y_c$, at which $U(y_c)=c$.
At the height $y_c$, called the critical height, 
equation (\ref{TG}) has a regular singular point.
Many of the features of the unstable modes are related to the presence
of this singular point.
Equation (\ref{TG}) together with the boundary conditions $\phi \to
0$ for $y\to \pm \infty$, forms an eigenvalue problem for the
complex eigenvalue $c$.

The Taylor-Goldstein equation (\ref{TG}), has been studied for many different
density and velocity profiles and
is known to have four different classes of modes as solutions \cite{Banks76}:
(i) For some conditions unstable modes exist with the real part of the phase velocity
within the range of $U$.
The phase speed of these modes satisfies Howard's semi-circle theorem
$|c-1/2 (\sup\{U\}+\inf\{U\}) |<1/2\,|\sup\{U\}-\inf\{U\}|$.
Furthermore, if these modes exist the Miles-Howard theorem \cite{Howard61} 
guarantees that somewhere in
the flow the local Richardson number defined by:
\begin{equation}
\Ri(y)=\frac{J(y)}{[U'(y)]^{\,2}}
\end{equation}
must be smaller than 1/4.
As discussed in the introduction,
for some velocity and density profiles, the class of the unstable modes
can be further divided
into two subclasses of unstable modes, those whose phase velocity is zero with respect the
mean flow (Kelvin-Helmholtz modes) and those whose phase velocity is non-zero,
the Holmboe modes.
(ii) The complex conjugates of the unstable modes constitute the
second class of
modes. That is, for each ($c_u,\phi_u$) that is a solution
of (\ref{TG}) that describes an unstable mode 
there is a solution ($c_d,\phi_d$) that describes a
damped mode with $c_d=c_u^*$ and  $\phi_d=\phi_u^*$. 
(iii) The
third class of modes consists of non-singular traveling modes
with phase velocity outside the range of $U$. These are internal
gravity wave modes modified by the shear.
(iv) Finally, due to the singularity at the critical height $y_c$ a
continuum of singular neutral modes exist with a singular behavior
of the first derivative of the steam function at the critical
height. The phase velocity of these modes lies within the range of
$U$. For these modes an asymptotic analysis close to the critical
height shows that the stream function can be written as the sum of
two linearly independent solutions:
\begin{equation}
\label{ASC} \phi \simeq A^{\pm} (|y-y_c|^{s_a}+\dots) \,\, +
\,\, B^{\pm}(|y-y_c|^{s_b}+\dots)
\end{equation}
where $s_{a}=1/2-\sqrt{1/4-\Ri(y_c)}$,
$s_{b}=1/2+\sqrt{1/4-\Ri(y_c)}$ and the index $\pm$ corresponds to
the solution above or below the critical height. 
The solutions in equation (\ref{ASC}) are known as the Frobenius solutions. 
If these modes are
marginally unstable the Frobenius solutions need
to be considered as the limit $\Ima\{c\}\to 0^+$ and yield the connection
formulas: 
\begin{equation}
\label{CF}
A^-=e^{is_a\pi}A^+ \,\, {\mathrm{and}} \, \, \, B^-=e^{is_b\pi}B^+ \, .
\end{equation}   
For each wavenumber and fixed density and velocity profiles more than
one of the above mentioned modes can exist. Consequently, it is interesting to
try and find boundaries, if they exist, that separate these four
classes of modes. This is what we try to achieve in what follows
for a specific example of shear and stratification.

The example we investigate follows Hazel's \cite{Hazel72} work
where the velocity profile is given by:
\begin{equation}
\label{u}
 U(y)=\tanh(y).
\end{equation}
Note that the problem has been is nondimensionalized using the
half-thickness and the half velocity change of the shear layer.
For the stratification as, in \cite{Hazel72}, we pick the
squared Brunt-V\"ais\"ala frequency's functional form (in nondimensional units) 
to be given by:
\begin{equation}
\label{J}
J(y)=J_0 \, \sech^2( R y)
\end{equation}
where $R^{-1}$ gives the non-dimensional length scale of the
density stratification. The symmetry $J(y)=J(-y)$ and
$U(y)=-U(-y)$ implies that if ($c,\phi(y)$) is the eigenvalue and
eigenfunction of an unstable mode, then ($-c^*,\phi(-y)$) is also a
solution of equation (\ref{TG}) that describes an unstable mode
traveling in the opposite direction. This allows us with no loss
of generality to focus only on the modes that have $Re\{c\} >0$.

\begin{figure}
\includegraphics[width=8cm]{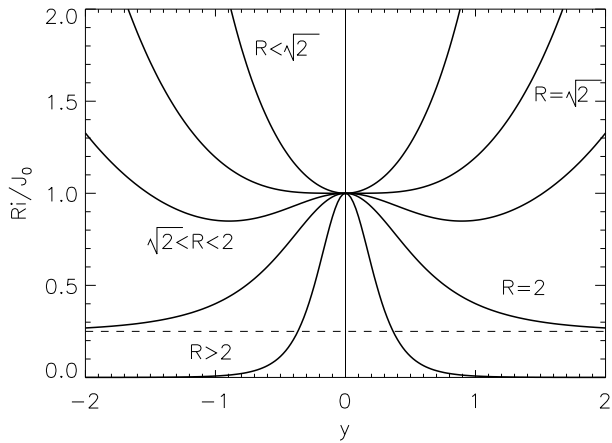}
\caption{\label{fig1} The local Richardson number $Ri(y)$ normalized by $J_0$
for  different values of the parameter $R$}
\end{figure}
The following observations illuminate the origin and possible location
of the stability boundaries.
First, as noted by Hazel \cite{Hazel72}, 
for the density and velocity profile given in equations (\ref{J},\ref{u})
respectively, the functional form of the local Richardson number crucially
depends on the value of the parameter $R$ (see figure \ref{fig1}). 
For $R \le  \sqrt{2}$, the local Richardson number
has a unique minimum at $y=0$ given by $\Ri(0)=J_0$.
This implies that we can only have instability if $J_0<1/4$ \cite{Howard61} 
and any stability boundary must lie within this region or on its boundary.
For $\sqrt{2}<R<2$ two minima exist
symmetrically around zero and therefore we can have instability 
for values of $J_0$ larger than
$1/4$.
For $R = 2$, $\Ri(y)$ obtains its minimum value for
$y=\pm \infty$ with $\Ri(\pm \infty) =1/4J_0$ and therefore instability can occur only if
$J_0<1$. Finally for larger values of
$R$, $\Ri(y)$ decays exponentially to zero as $y$ becomes large
and an instability might be present at arbitrarily large values of $J_0$.

\begin{figure}
\includegraphics[width=8cm]{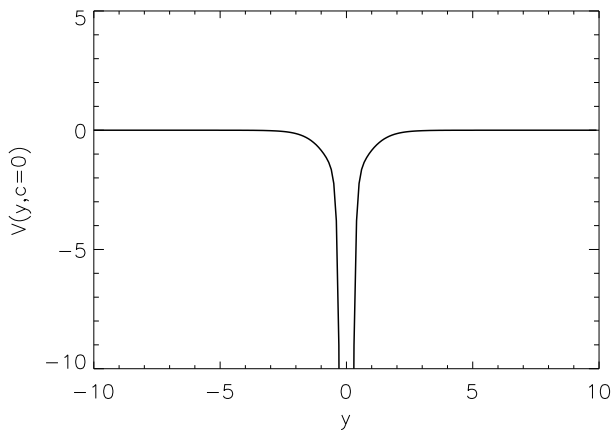}
\caption{\label{fig2} The potential $V(y,c)$ in equation (\ref{SCH}) for c=0}
\end{figure}

In order to make the second observation 
we need to transform equation (\ref{TG}) in to
an alternative form.
Denoting $E=-k^2$ and $V(y,c)=U''/(U-c)-J/(U-c)^2$ we can rewrite
the Taylor-Goldstein equation as:
\begin{equation}
\label{SCH}
\phi''=[V(y,c)-E]\phi \, .
\end{equation}
Now, if we invert the problem and ask the question, for
a given real phase velocity $c$ what is the value of $k^2=-E$ that
satisfies the above equation and boundary conditions, we end up with a Schr\"odinger
problem for a particle in a potential well given by $V(y,c)$.
Figure \ref{fig2} shows the resulting singular potential for $c=0$,
$R=1$ and $J_0=2$.
The solution of this particular problem for $R=1$ gives the stability boundary for the
unstable modes given by the relation $J_0=k(1-k)$ as shown in \cite{Miles63}.
\begin{figure}
\includegraphics[width=8cm]{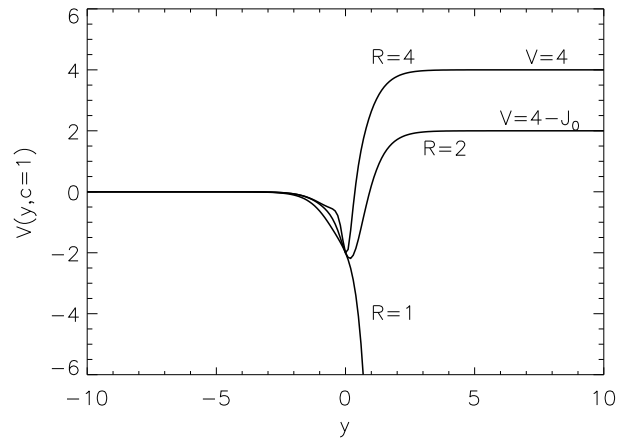}
\caption{\label{fig3}The potential $V(y,c)$ in (\ref{SCH}) for c=1 for three different
values of $R$. Only when $R \ge 2$ bounded states are allowed.}
\end{figure} 
Figure \ref{fig3} shows the resulting potential for $c=1$ and
three values of $R=1,2,4$. If $R < 2$ no solution exists
that satisfies the boundary conditions since this problem
corresponds to finding bounded states in an unbounded potential
well ($V(y,1) \to -\infty$ for $y \to +\infty$). However if
$R \ge 2$ we have to solve for the eigenstates/energy levels
of a particle in a finite potential well. For this case a discrete
number of bounded energy states with ``energy levels" $E_n$ exist.
Each state corresponds to a mode with $c=1$ and wavenumber $k_n=\sqrt{-E_n}$.
Since the phase speed of free gravity waves is a decreasing
function of wavenumber, we expect that wavenumbers
smaller than $\sqrt{-E_n}$ will have $|c|<1$ and could
possibly be unstable. Of course, whether these solutions indeed
represent marginally unstable modes still needs to be
demonstrated. Note that marginally unstable modes
with phase velocity equal to the maximum velocity of the shear flow 
have been reported before in the literature in interfacial gravity wave generation problems
\cite{Alexakis02,Alexakis04,Churilov05}.

Besides the modes with $c=1$, 
other neutral modes that may be marginally
unstable should have $c$ in the range of $U$ and therefore
must be singular. These modes can be written in terms of the asymptotic
expansion in equation (\ref{ASC}). 
For example let $\phi(y)=A^+f_a(y)+B^+f_b(y)$ be the exponentially decaying solution of
of the Taylor-Goldstein equation \ref{TG} for $y>y_c$
(where $f_a$ and $f_b$ are the two Frobenius solutions in \ref{ASC}).
$A^+$ and $B^+$,  with no loss of generality, are assumed real.
The connection formulas for marginally unstable modes \ref{ASC} imply that
below the critical layer ($y<y_c$) the solution will be 
$\phi(y)= [A^+ \cos(s_a \pi) f_a(y) + B^+ \cos(s_b \pi) f_b(y)]$
$+       i[A^+ \sin(s_a \pi) f_a(y) + B^+ \sin(s_b \pi) f_b(y)].$
However, only one linear combination of $f_a$ and $f_b$
will give an exponentially decaying solution for $y\to -\infty$.
Therefore, if $A^+$ and $B^+$ are such that the real part of $\phi(y)$
is decreasing exponentially to zero for $y\to -\infty$ the imaginary part of
$\phi(y)$ will increase exponentially.
Since both the real and the imaginary parts
of $\phi$ need to satisfy the boundary conditions at infinity
it seems unlikely
that such a solution exists unless one of the two coefficients
$A^+$ or $B^+$ is zero. Therefore, a possible  marginally unstable mode would
be a singular mode that is proportional to only one of the two
independent Frobenius  solutions in equation (\ref{ASC}) (i.e. $A^{\pm}=0$ or
$B^{\pm}=0$). However the existence of such a solution is not guaranteed,
so in addition, a suitable value of $k^2$ and $c<1$ needs
to be found (if it exists) so that both boundary conditions are
satisfied. We note that the case in which the  $c=0$ modes form a stability
boundary corresponds to a special
case of this class of marginally unstable  modes.

\section{Numerical Methods}
\label{NumMeth}

The strategy that we adopt in order to find the unstable regions
is the following. First we find the location of the neutral modes
described in the previous section and then investigate if these
modes are indeed marginally unstable by solving for the growth rate
(or for $\Ima\{c\}$) in the neighborhood of these modes.
Four different numerical codes were used.
Each code integrates the Taylor-Goldstein equation (\ref{TG}) 
using a fourth order Runge-Kutta method 
(tested for different resolutions to verify convergence)
and a shooting method is used to determine
the eigenvalue in the four different eigenvalue problems 
that we describe in more detail below \cite{NumericalMethods}.

In the first eigenvalue problem we take $c=0$.
For a given value of $J_0$ we integrate the Taylor-Goldstein
equation from zero to infinity with initial conditions given by
one of the two Frobenius solutions (\ref{ASC}), and look for
the eigenvalue $k^2$ such that the boundary condition at
infinity is satisfied.
Note that for symmetry reasons, when c=0, only the positive
$y$-axis needs to be considered.
The solution of this problem provides us
with the stability boundary for the Kelvin-Helmholtz modes, as shown in \cite{Hazel72}.
We will refer to this problem as eigenvalue problem one. The
values of $J_0$ that satisfy the condition $c=0$ for a wavenumber
$k$ will be denoted by the curve  $J_0=J_{KH}(k)$.

For the second eigenvalue problem we take $c=1$
and solve the Schr\"odinger problem for the potential $V(y,1)$ 
by integrating equation (\ref{SCH}) from $-`\infty$' to $+`\infty$' and look
for the eigenvalue $E_n$ for which the boundary condition at $+`\infty$' is satisfied
(where by $`\infty$' we mean sufficiently large value of $y$).
The modes evaluated from this process are non-singular neutral modes,
and form a curve in the $(J_0-k)$ plane that we denote by $J_0=J_1(k)$.

In the third problem we are looking for the values of
$c$ ($c\ne0$ and $\Ima\{c\}=0$) and $k$ such that one of the two independent 
Frobenius solutions of equation (\ref{ASC})
satisfies both boundary conditions at infinity.
In more detail the third code begins with an initial ``guess"
of both $c$ and $k$ and integrates from the critical height
both forward and backward up to some sufficiently large positive/negative 
value of $y$.
The initial conditions for $\phi$ and $\phi'$ 
at $y=y_c\pm dy$ (where $dy$ is a small deviation from the critical height) 
are obtained from
equation (\ref{ASC}) with either $A^\pm=0$ or $B^\pm=0$
(both cases were tested) and terms up to second order in
the expansion are used. As argued in the previous section,
only the solution that it is proportional to one
of the two linearly independent solutions of (\ref{ASC})
can satisfy the connection formulas across the critical
layer (for marginally unstable singular modes) 
and both boundary conditions at infinity.
Any other linear combination of the Frobenius solutions (\ref{ASC})
will lead to a singular neutral mode that does not belong
to an instability boundary. 
Since we begin with either $A^\pm=0$ or $B^\pm=0$ we are
only left with two conditions to satisfy at $\pm `\infty$'
and two parameters to change ($c$ and $k$). With
a bisection method \cite{NumericalMethods} the code converged to a single
pair of values of $c$ and $k$. 
For one of the two initial conditions
(say $A^\pm=0$) the code converged to the $c=1$ solution,
and for the second initial condition (say $B^\pm=0$) the code
converged to a value of $c$ and $k$ with $-1< c < 1$.
The solutions with $c\ne1$ comprise a class of
singular neutral modes forming a curve in the $(J_0-k)$ plane
that we will denote by $J_0=J_{1S}(k)$.

However, the existence of the above mentioned neutral modes does
not guarantee that they form a stability boundary.  Hence,
a code that solves the full eigenvalue problem for the complex
eigenvalue $c$ was used to verify that the above mentioned
neutral solutions constitute the stability boundaries.
This fourth code integrates the Taylor Goldstein equation from $\pm`\infty$'
to zero for some given values of $J_0$ and $k$ 
and a Newton-Raphson method is used to find the value of $c$ 
for which the two solutions match at zero.
No approximation is used to cross the critical height and  
for this reason this code converged only for unstable modes (Im$\{c\}\ne0$)
and neutral non-singular gravity waves ($|c|>1$). 
Note that this code encounters problems with convergence when the examined
parameters are very close to a neutral singular mode stability boundary
because of the almost singular behavior in the neighborhood of the critical height
that needs to be resolved.

\section{Results}
\label{Res}
\subsection{Small values of $J_0$}
\label{SmJ}

We begin with relatively small values of $J_0$ 
(order unity or less)
which is the case that has been
previously examined by \cite{Hazel72} and \cite{Smyth89}. However in
\cite{Hazel72} and \cite{Smyth89} the authors determined the instability boundaries in the
$(J_0-k)$ plane by solving the full eigenvalue problem in the
examined parameter space, finding in that way the regions with non-zero
growth rates. This is a very difficult task since 
in order to determine the
instability region,
the entire $(J_0-k)$ plane needs to be mapped for every value of $R$. 
Here we use a different approach.
We find the location of the neutral modes corresponding to the
different eigenvalue problems described in the last
section and show that these form instability boundaries
by solving the complex value of the phase velocity only for a few
values of the global Richardson number.
\begin{figure}
\includegraphics[width=8cm]{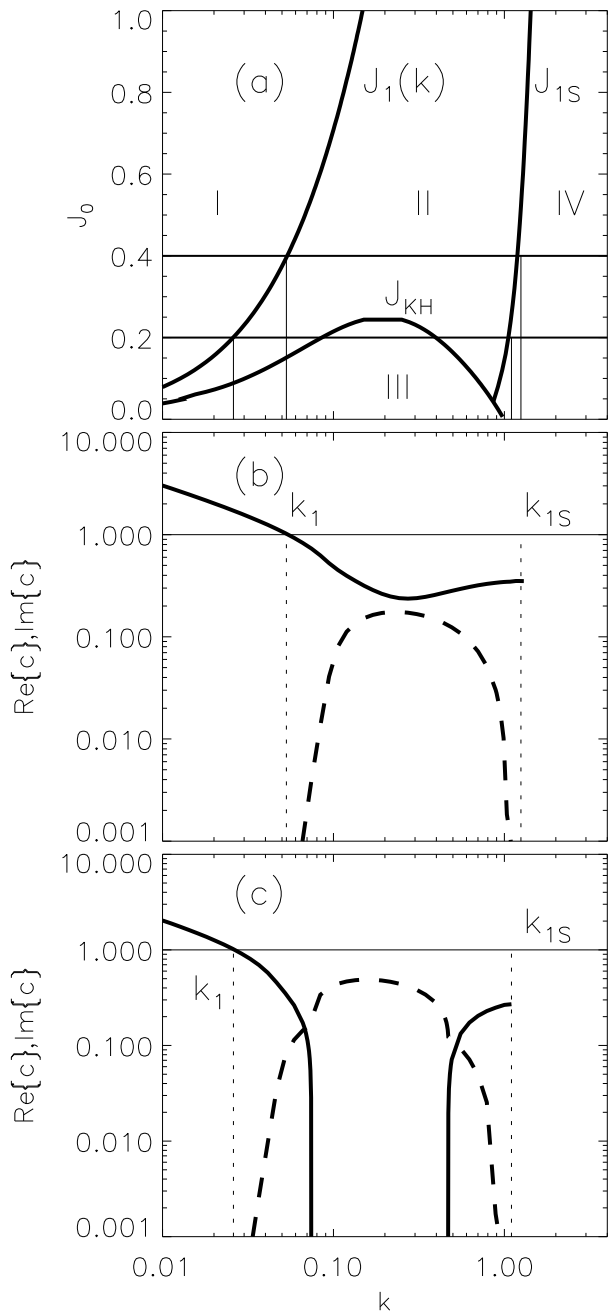}
\caption{\label{fig4} Panel (a): The stability boundaries for $R=4$.
In region  (I) stable gravity waves exist with $|c|>1$.
In region  (II) unstable waves with
$0<|$Re$\{c\}|<1$ exist.
In region (III) unstable KH-modes with Re$\{c\}=0$ exist.
In region (IV) only neutral singular modes exist.
The stability boundary $J_{1}(k)$
is composed of the modes with $c=1$.
The stability boundary $J_{KH}(k)$
is composed of the modes with $c=0$.
The stability boundary $J_{1S}(k)$
is composed of the singular neutral modes with $c\ne 0$.
The horizontal lines correspond to the values of $J_0$
examined in panels (b,c).
Panel (b): The real (solid line) and imaginary (dashed line) phase velocity for $J_0=0.4$.
Unstable modes exist for wavenumbers in the range $k_1<k<k_{1S}$.
Panel (c): The real (solid line) and imaginary (dashed line) phase velocity for $J_0=0.2$.
Unstable modes exist for wavenumbers in the range  $k_1<k<k_{1S}$.
In both cases Im$\{c\}$ is approaching zero in the boundaries of the range $(k_1,k_{1S})$.
All panels share a common $x$-axis.
}
\end{figure}

In figure \ref{fig4} we show the results for the case $R=4$.
As was found in previous work \cite{Hazel72,Smyth89} 
the $(J_0-k)$ plane can be divided in four regions
(see panel (a)). In the first  region (I)  free gravity waves exist
with real phase velocity of absolute value greater than one. 
In the second region (II) unstable Holmboe
waves exist with the real part of the phase velocity smaller than one but different from zero.
In the third region (III) unstable Kelvin-Helmholtz modes are present
with the real part of the phase velocity equal to zero. Finally in the fourth region (IV) only
singular neutral modes exist. The three lines that separate these regions were constructed
by finding the neutral modes with the properties described in the previous section.
In more detail
the curve $J_1(k)$ that separates region (I) from region (II) is composed of modes that
have phase velocity equal to one ($c=1$). The curve $J_{KH}(k)$ 
that separates region (II) from region (III)
is composed of singular modes with $c=0$. Finally the curve $J_{1S}(k)$ that separates
region (II) from region (IV) is composed of singular modes with $c\ne 0$.

Further illumination of how the properties of a mode change as we increase the
wavenumber is gained by looking at panels (b,c) of figure \ref{fig4} that show
the dispersion relation (phase velocity as a function of the wavenumber)
for two values of $J_0$ (panel (b) $J_0=0.4$ and panel (c) $J_0=0.2$). 
A logarithmic scale is used to focus on the details of small wave-numbers
and small phase velocities.
For the
$J_0=0.4$ case, for small wavenumbers the phase velocity is real and decreases
with $k$ until the phase velocity becomes equal to one ($c=1$ mode)
for some wavenumber $k_1\simeq 0.05$ such that $J_1(k_1)=0.4$.
For wavenumbers larger than $k_1$ the phase
velocity becomes complex indicating that the mode with $c=1$ belongs to the
stability boundary.
As we further increase the wavenumber there is a critical value
$k_{1S}$ for which the imaginary part of the phase velocity becomes
zero again, with the real part of the phase velocity remaining
finite and smaller than one. This mode with
wavenumber $k_{1S}$ can only be a singular neutral mode and can be
represented by the expansion given in equation (\ref{ASC})
with the connection formulas given in equation (\ref{CF}). 
Indeed the
difference between the wavenumber where the imaginary part of $c$
becomes zero (found by the fourth eigenvalue problem) and the
wavenumber $k_{1S}$ (found by solving the third eigenvalue problem), 
is found (for this case
and all the cases examined) to be of the order of the numerical
error. 
The small difference is attributed to the difficulty in
resolving the almost singular region around $y_c$ with the fourth code
when $\Ima\{c\}$ is very small.

For smaller values of the global Richardson number ($\Ri(0)=J_0<0.25$)
we have the extra feature that
as we approach the stability boundary composed of modes with $c=0$
the real part of the phase velocity decreases to zero. It remains zero
in region (III) and then starts to increase again in region (II) until
region (IV) is met. We note that in region (IV) a continuum of singular stable modes exist
and for this reason we do not plot the phase velocity after this point.

\begin{figure}
\includegraphics[width=8cm]{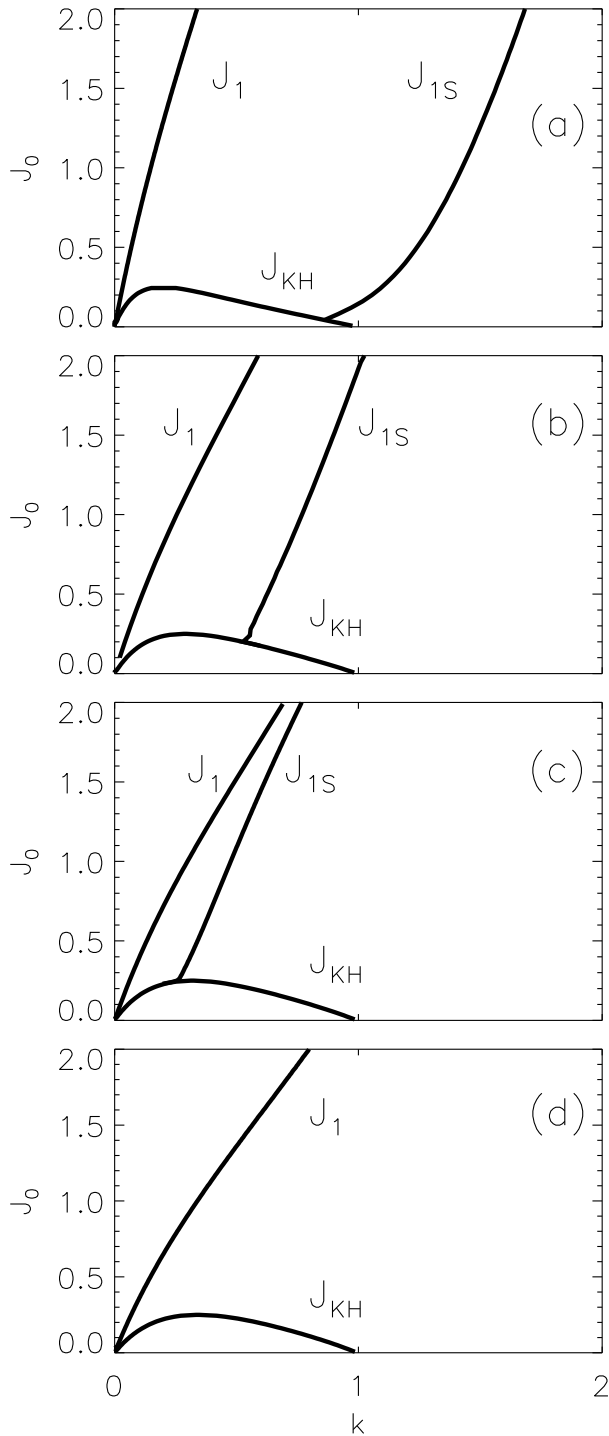}
\caption{\label{fig5} The dependence of the instability boundaries on the parameter $R$.
The four panels show the stability boundaries for four values of $R$.
Panel (a) $R=4$, panel (b) $R=2.5$, panel (c) $R=2.2$, panel (d) $R=2$.
Note that for the last case, $R=2$, in panel (d) the stability boundary $J_1(k)$
composed of modes with the property $c=1$ and the boundary with singular modes
$J_{1S}(k)$ have
collapsed together and the resulting boundary
now defines the region that non-singular neutral gravity waves exist.
All panels share a common $x$-axis.}
\end{figure}

The next question we examine 
is how the stability boundaries change as we vary
the value of the parameter $R$. In figure \ref{fig5} we plot the stability boundaries
for four different values of $R$
(panel (a) $R=4$,
 panel (b) $R=2.5$,
 panel (c) $R=2.2$ and
 panel (d) $R=2$).
For large values of the parameter $R$ ($R>2$) the
$(J_0-k)$ plane is divided in four regions determined by the
boundaries $J_1,J_{1S},J_{KH}$ as discussed in more detail in
figure \ref{fig4}.
The wavenumbers $k$ of the Holmboe unstable modes are confined
in a finite strip. 
As we decrease the value of $R$ the
width of the strip with Holmboe unstable wavenumbers becomes
smaller (see panel b and c). When $R=2$ (panel d) the two
boundaries $J_{1S}$ and $J_{1S}$ collapse together. In this case
$J_1$,$J_{1S}$ does not provide a stability boundary but continues to separate
the region that non-singular gravity wave modes exist (on
the left of $J_1$) from the region that only singular neutral modes exist
with $0<c<1$ (on the right of $J_1$). The only unstable modes for this case are the
Kelvin-Helmholtz modes confined in the finite region in the left bottom
corner.

\begin{figure}
\includegraphics[width=8cm]{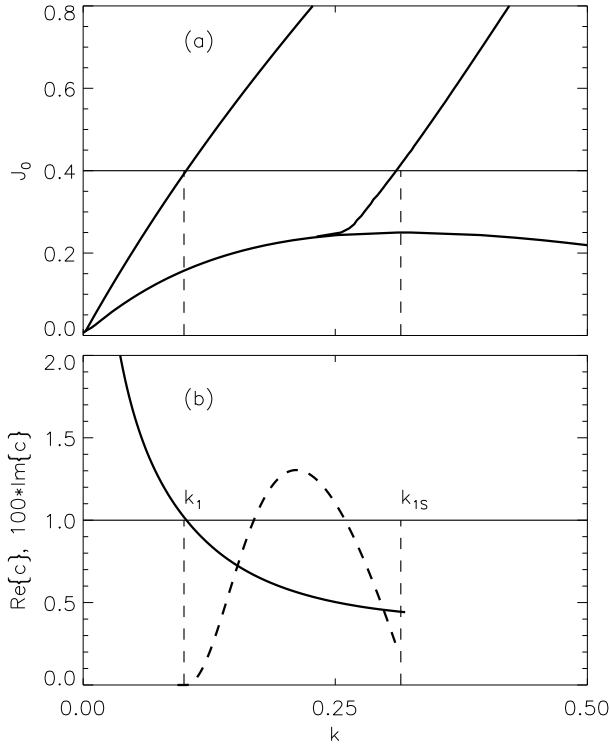}
\caption{\label{fig6} The stability boundaries panel (a) and the dispersion relation
($J_0=0.4$) panel (b)
for the case that $R=2.2$.
In panel (b) solid line gives the real part of $c$
and the dashed line gives the imaginary part of $c$.
Unstable Holmboe modes exist for this case also but are
confined in a very narrow region.}
\end{figure}

We note here that \citep{Smyth89} did not find Holmboe instability for values of
of $R$ smaller than 2.4. This is probably because the width of instability strip
becomes very small for values of $R$ smaller than 2.4.
Furthermore, close to the stability boundaries the growth rate 
is very small and as a result it is even harder to find a positive growth rate
using the method used in \citep{Smyth89} when the two stability boundaries are
too close.
To verify that this strip is unstable we plot in figure \ref{fig6} a close up
of the instability region and the dispersion relation for $J_0=0.4$ 
for the case of $R=2.2$.

\subsection{Higher harmonics}
\label{HH}

As we have shown, finding the modes that satisfy $c=1$ can be reduced to 
finding the eigenstates of a particle in a finite potential well.
It is well known from quantum mechanics that if a potential well is deep enough
a finite number $N\ge 1$ of bound states will exist with $N$ different energy levels $E_n$
and wavefunctions $\phi_n$ ($n=1,2,\dots,N$).
If the modes with $c=1$
form one of the stability boundaries for the Holmboe instability, it is
natural to ask what happens for the case that $J_0$ is large enough
(i.e. the potential well in the Schr\"odinger problem (\ref{SCH}) is deep enough) 
so that more than one mode satisfies the condition $c=1$ 
( i.e. more than one bounded eigenstate is allowed).
\footnote{We note that a discontinuous density profile will lead
to a delta-function behavior of $J(y)$ that allows only one eigen-state
and for this reason higher harmonic waves are not present in step-function
density profiles like the ones examined in \cite{Holmboe62}. }

To investigate this we solve for the modes with $c=1$
for values of $J_0$ up to $10^2$.
We find that the higher harmonics of the gravity waves with $c=1$ indeed
exist and correspond to new stability boundaries that we denote as $J_n(k)$.
The second harmonic ($n=2$) with $c=1$ appears for $J_0\simeq 28$
and the third harmonic ($n=3$) for $J_0\simeq 80$.
We also find that 
for each $c=1$ mode 
a marginally singular neutral mode with $0<c<1$ also exists.
These modes form new curves in the $(J_0-k)$ plane given by $J_{nS}(k)$.
These curves $J_n(k)$ and $J_{nS}$, define new stripes of unstable regions.

\begin{figure}
\includegraphics[width=8cm]{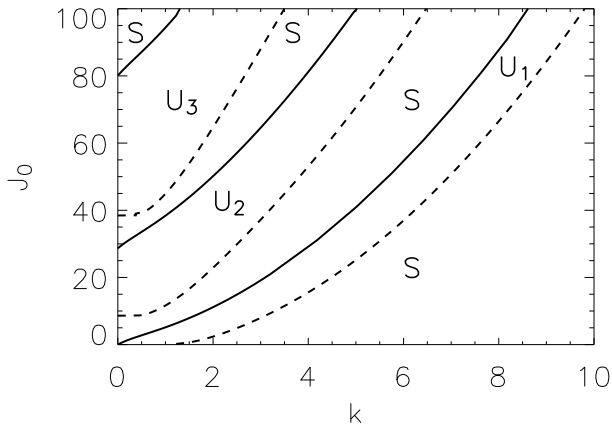}
\caption{\label{fig7} Instability diagram for larger values of $J_0$
and $R=4$.
Solid lines show $J_n(k)$, dashed lines show $J_{nS}(k)$.
The stable regions are marked with ${ S}$. Regions marked with ${ U_n}$
are regions that the $n$-th harmonic becomes unstable. }
\end{figure}

\begin{figure}
\includegraphics[width=8cm]{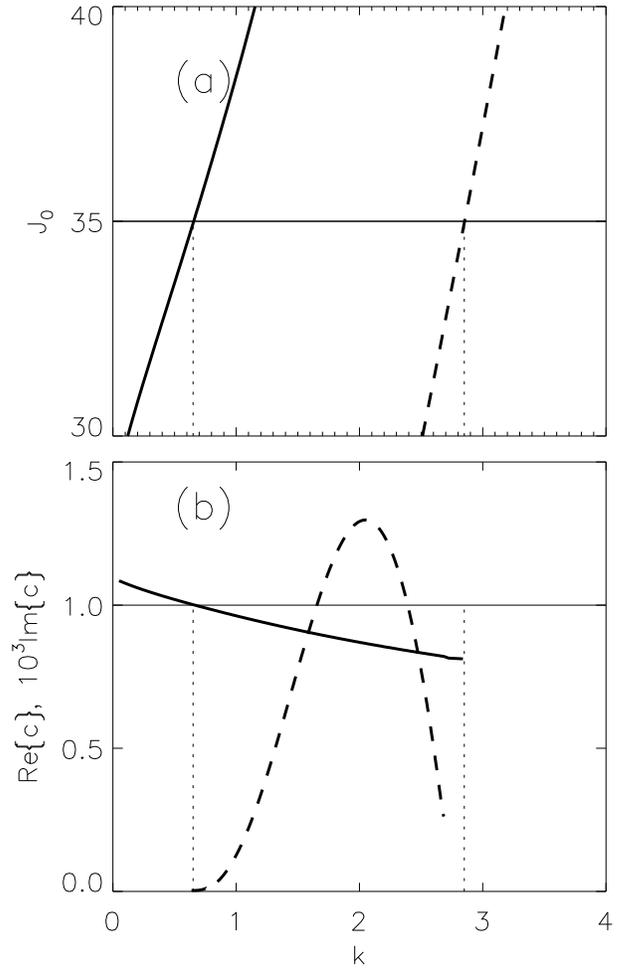}
\caption{\label{fig8}
Panel (a):The stability boundaries for the second harmonic
for $R=4$. Solid line gives $J_2(k)$, dashed line gives $J_{2S}(k)$.
Panel (b): The real part of the phase velocity (solid line)
and the imaginary part of the phase velocity (dashed line) for $J_0=35$.}
\end{figure}

\begin{figure}
\includegraphics[width=8cm]{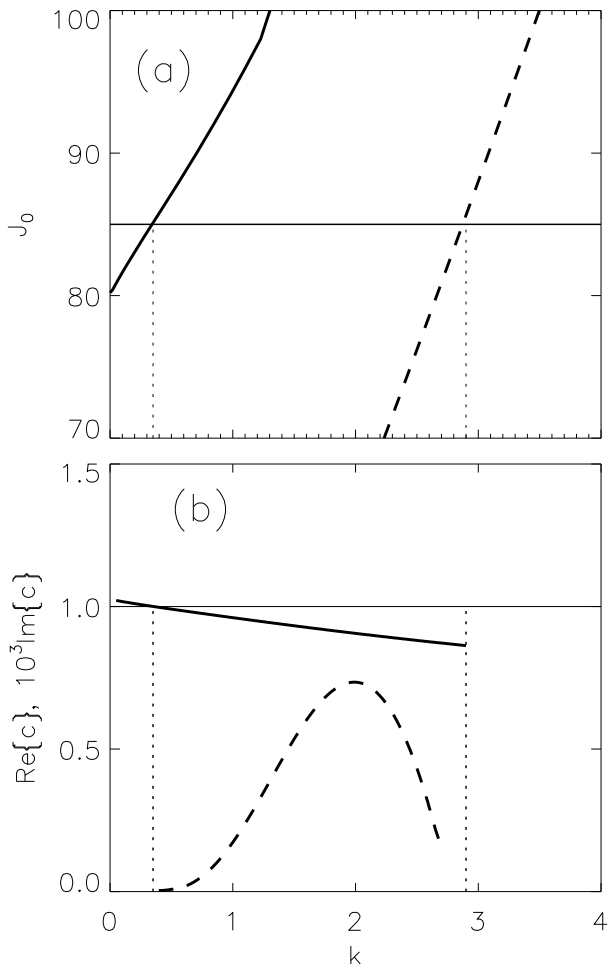}
\caption{\label{fig9} Panel (a). The stability boundaries for the third harmonic
for $R=4$. Solid line shows $J_3(k)$ and dashed line shows $J_{3S}(k)$.
Panel(b) The real part of the phase velocity (solid line)
and the Imaginary part of the phase velocity (dashed line) for $J_0=85$.}
\end{figure}

In figure \ref{fig7} we plot the stability domain for values of $J_0$
up  100 for $R=4$. The stable regions are marked with $S$ and the regions that the
$n$-th harmonic becomes unstable are marked with $U_n$. 
The solid  lines show $J_n(k)$    and correspond to modes with $c=1$ and 
the dashed lines show $J_{nS}(k)$ and correspond to marginally unstable singular modes. 
As before we verify that unstable modes exist in the regions between $J_n(k)$ and $J_{nS}(k)$ 
by solving for the complex phase velocity of the modes in these regions. In
figures \ref{fig8} and \ref{fig9} (panel (a)) we show a close up of the stability boundaries
$J_2(k),J_{2S}(k)$ and $J_3(k),J_{3S}(k)$ respectively,
and the dispersion relation in panel (b). Figure \ref{fig8}  and \ref{fig9} demonstrate
how the second and the third gravity wave harmonic become unstable.

\begin{figure}
\includegraphics[width=8cm]{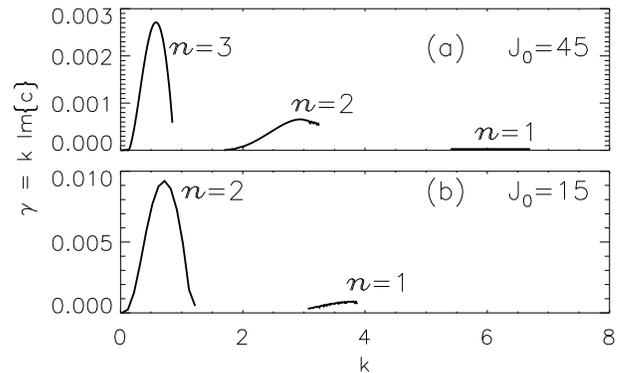}
\caption{\label{fig10} The growth rate for the different unstable harmonics.
Panel (a) $J_0=45$, $R=4$.
Panel(b)  $J_0=15$, $R=4$.
For the smallest harmonics for which
the Im$\{c\}$ was very small the code was able to
find the growth rate only around the peak and not
close to the stability boundaries.
For the $J_0=45$ case in panel (a) the growth rate of the first harmonic ($n=1$)
was too small for the code to be able to resolve
the critical height and the dark line just indicates
the location of the instability region based on the stability boundaries
shown in figure \ref{fig7}.}
\end{figure}

We note finally that for a fixed value of $J_0$ the highest
harmonic has the largest growth rate.
In figure \ref{fig10} we show the growth rate $\gamma=k$Im$\{c\}$ as
a function of the wavenumber for $R=4$ and two different values of
$J_0$. The growth rate is significantly larger for the highest
harmonic. This makes the higher unstable
harmonics possible to be detected experimentally, since it is
the most unstable modes that are usually observed. 

\section{Conclusions and physical description}
\label{Cons}

In this paper we investigated Holmboe's instability in stratified
shear flows for smooth density and velocity profiles. Using a
specific model of the density and velocity profiles
\cite{Hazel72,Smyth89}, we were able to determine the instability
regions for large values of the global Richardson number including
regions not previously noted in the literature.

Focusing on the nature of the stability boundaries
that enclose the Holmboe unstable modes, we have shown that for
moderate values of $J_0$ and when the density stratification
length scale is sufficiently smaller than the shear length scale
(i.e. $R > 2$) the ($J_0-k$) plane is divided into four
regions: 
(I) a region where neutral gravity waves exist (i.e. modes
with real phase velocity outside the range of the velocity shear);
(II) a region where unstable traveling waves exist (i.e. Holmboe
modes); 
(III) a region where unstable
modes with the real part of the phase velocity being zero exist (i.e. Kelvin Helmholtz modes); 
(IV) and finally a region where only singular neutral modes exist. 
We determined the modes that comprise the boundaries and separate these
four classes of modes.
For small $J_0$ ($J_0<1/4$) Kelvin-Helmholtz modes exist. They
are restricted in a bounded domain enclosed by the boundary
$J_0=J_{KH}(k)$ determined by the modes with zero phase velocity.
For larger $J_0$, a strip of Holmboe unstable modes exist. This
instability strip is determined by the two boundaries
$J_0=J_1(k)$ and $J_0=J_{1S}(k)$.
In more detail, we have demonstrated that for a given value of the
global Richardson number $J_0$ the unstable modes have wavenumbers 
$k$ that lie in the range $k_1<k<k_{1S}$ where $k_1$ is
the wavenumber of the mode that has phase velocity equal to one
and satisfies $J_0=J_1(k_1)$ and $k_{1S}$ is the wavenumber of a
singular marginally unstable neutral mode. $k_{1S}$ satisfies
$J_0=J_{1S}(k_{1S})$.
%
We have shown how the value of $k_1$ and $k_{1S}$  can be
determined to desired accuracy 
(and therefore how to construct the boundaries $J_1(k)$ and $J_{1S}(k)$)
by solving a Schr\"odinger
eigenvalue problem for a particle in a potential well.

For large values of the global Richardson number  
more than one strip of instability may appear. 
Each new instability strip is  
confined between the pair of curves $J_0=J_n(k)$ and $J_0=J_{nS}(k)$.
As before $J_n(k)$ can be determined for a given value of $J_0$ 
by finding the wavenumber $k_n$ of the $n-$th 
harmonic of the gravity wave spectrum with $c=1$ and 
$J_{nS}(k)$ is determined by finding the wavenumber $k_{nS}$
of the singular marginally unstable mode that is met first as
we increase the wavenumber from $k_n$.

%

The results of this paper show a deep connection between the free gravity wave spectrum 
and the Holmboe unstable waves. The overall picture looks as follows. 
For large stratification and small wavenumbers the free 
gravity wave spectrum is only slightly modified by the shear
and is composed of a discrete finite number $N$ ($N \ge1$) of stable modes
($\phi_n,c_n$) with $n=1,2,\dots,N$.
As the wave
number is increased the phase speed decreases roughly as the square root
of the wavenumber. As the phase speed approaches the maximum shear velocity 
($\sup\{U\}$),
the functional form  of the stratification plays a crucial role. 
If the stratification length scale is larger than a critical value 
(more precisely if $\lim_{y\to \infty}\Ri(y)\to \infty$) 
the phase speed approaches $\sup\{U\}$ asymptotically as $k\to \infty$,
always remaining outside the range of $U$. 
In this case the only unstable wavenumbers are the Kelvin-Helmholtz modes
that have zero phase velocity and appear only for a finite range of
the global Richardson number.
If the stratification length scale
is smaller than this critical value
(more precisely if $\lim_{y \to \infty}\Ri(y)\to 0$),
the phase speed of the waves reaches the value of the maximum wind speed
for a finite value of the wavenumber $k=k_n$. 
As we increase the wavenumber further, the 
modes become unstable with the real part of the phase velocity smaller than one.
The instability persists up to another critical value of the wavenumber
$k=k_{nS}$. The mode with this wavenumber exhibits a singular behavior
at the critical height. For wavenumbers smaller than $k_{nS}$ a continuum
of singular neutral modes exist. 

The findings in this work suggest several possibilities for further
investigation. An important aspect that needs to be investigated
is the effect of viscosity and diffusivity. As the work in \cite{Nishida87,Smyth90} 
has shown, the presence of finite viscosity and diffusivity decreases 
the growth rate and this effect is expected to be stronger 
for the higher harmonics. This may limit
attempts to detect the higher harmonic modes 
via numerical and experimental studies. 
Therefore, an investigation of the viscosity effects is further needed
to find the critical Reynolds number for which the higher modes become
unstable and this should be the next step in the study of the higher harmonics
instability. 
Nonetheless, we note that in most astrophysical and geophysical flows, 
where Holmboe's instability is of great importance,
the Reynolds and Peclet numbers are large enough that the inviscid flow
examined here is a good approximation, and depending on the stratification the higher
harmonic modes might dominate.

Finally, besides viscosity and dissipation, three dimensional effects, nonlinear
evolution and saturation, as well as physical interpretation of these results 
through simplified layer models are all important issues that we plan to
investigate in our future work.

\begin{acknowledgments}
The author acknowledges support from NCAR/GTP.
The author is grateful to L. Howard for his suggestions and references
at the beginning of this work, and would also like to thank 
K. MacGregor, J. Mason and J. Sukhatme for their comments. Finally, during the review
process the comments by W.D. Smyth helped to bring this
manuscript to its final form and his help is appreciated.
\end{acknowledgments}



\begin{thebibliography}{}



\bibitem[Rosner, Alexakis, Young, Truran \& Hillebrand (2002)]{Rosner01}
{ R. Rosner, A. Alexakis, Y. Young, J. Truran \&
W. Hillebrand}, ``On the C/O enrichment of novae ejecta".
{Astrophys. J.\/} {\bf 562}, L177--L179 (2002).

\bibitem[Armi \& Farmer (1988)]{Armi88}
{ L. Armi \& D.M. Farmer,}
``The flow of Mediterranean water through the Strait of Gibraltar".
{ Prog. Oceanogr. \/} {\bf 21}, 1--105 (1988).

\bibitem[Oguz, Ozsoy, Latif, Sur,  \& Unluata (1990)]{Oguz90}
{T. Oguz, E. Ozsoy, M.A. Latif, H.I. Sur \& U. Unluata,}
``Modeling of hydraulically controlled exchanged flow in the Bosphorous Strait".
{ J. Phys. Oceanogr. \/} {\bf 20}, 945--965 (1990).


\bibitem[Sargent \& Jirka (1987)]{Sargent87}
{ F.E. Sargent \& G.H. Jirka,}
``Experiments on saline wedge".
{ J. Hydraulic Eng. \/} {\bf 113}, 1307--1324 (1987).

\bibitem[Yoshida, Ohtani, Nishida \& Linden (1998)]{Yoshida98}
{ S. Yoshida, M. Ohtani, S. Nishida \& P.F. Linden,}
``Mixing processes in a highly stratified river".
{\it Physical Processes in Lakes and Oceans}, edited by J. Imberger
(American Geophysical Union, Washington DC, 1998)

\bibitem[Helmholtz(1868)]{Helmholtz68}
{ H. Helmholtz,} 
``Ueber discontinuirliche Fl\"{u}ssigkeitsbewegungen".
{ Wissenschaftliche Abhandlungen\/} {\bf 3}, 146 (1868).

\bibitem[Lord Kelvin(1910)]{Kelvin10}
{ Lord Kelvin},
``Influence of wind and capillarity on waves in water supposed frictionless".
{ Mathematical and Physical papers {\bf iv} Hydrodynamics and general Dynamics.\/} 76, (1910).

\bibitem[Howard (1963)]{Howard63}
{ L.N. Howard,}
``Neutral curves and stability boundaries in stratified shear flow".
{ J. Fluid Mech.\/} {\bf 16}, 333--342 (1963).


\bibitem[Howard \& Drazin (1966)]{Howard66}
{ L.N. Howard, P.G. Drazin,}
``Hydrodynamic stability of parallel flow of inviscid fluid".
{ Adv. in Appl. Mech.\/} {\bf 9}, 1--89 (1966).


\bibitem[Howard \& Maslowe (1973)]{Howard73}
{ L.N. Howard, \& S.A. Maslowe},
``Stability of Stratified Shear Flows".
{ Boundary Layer Meteorology.\/} {\bf 4}, 511--323 (1973).

\bibitem[ Drazin \& Reid (1981)]{Drazin81}
{P.G. Drazin \& W.H. Reid}
``{\it Hydrodynamic Stability"}
Cambridge University Press (1981)

\bibitem[Holmboe (1962)]{Holmboe62}
{J. Holmboe,}
``On the behavior of symmetric waves in stratified shear layers".
{ Geophys. Publ. \/} {\bf 24}, 67--113 (1962).

\bibitem[Caulfield (1994)]{Caulfield94}
{C.P. Caulfield,}
``Multiple linear instability of layered stratified shear flow".
{J. Fluid Mech. \/} {\bf 258}, 255--285 (1994).

\bibitem[Haigh \& Lawrence (1999)]{Haigh99}
{S.P. Haigh \& G.A. Lawrence,}
``Symmetric and non-symmetric Holmboe instabilities in an inviscid flow".
{ Phys. Fluids \/} {\bf 11}, 1459--1468 (1999).

\bibitem[Lawrence, Browand \& Redecopp (1991)]{Lawrence91}
{G.A. Lawrence, F.K. Browand \& L.G. Redecopp,}
``The stability of a sheared density interface".
{ Phys. Fluids-A.\/} {\bf 3}, 2360--2370 (1991).

\bibitem[Baines \& Mitsudera (1994)]{Baines94}
{P.G. Baines \& H. Mitsudera}
``On the mechanism of shear flow instabilities".
{ J. Fluid Mech. \/} {\bf 276}, 327--342 (1994).

\bibitem[Hazel (1972)]{Hazel72}
{S.P. Hazel,}
``Numerical studies of the stability of inviscid shear flows".
{ J. Fluid Mech. \/} {\bf 51}, 3261--3280 (1972).


\bibitem[Smyth \& Peltier (1989)]{Smyth89}
{ W.D. Smyth, \& W.R. Peltier,}
``The transition between Kelvin-Helmholtz and Holmboe instability;
An investigation of the over-reflection hypothesis".
{ J. Atmos. Sci. \/} {\bf 46}, 3698--3720 (1989).

\bibitem[Nishida \& Yoshida (1987)]{Nishida87}
{ S. Nishida \& S. Yoshida},
``Stability and eigenfunctions of disturbances in stratified two layer shear flow".
{\it Proc. Third Intl. Symp. on Stratified Flows, Pasadena, California, 3-5 February 1987} pp.28--34

\bibitem[Smyth, \& Peltier (1990)]{Smyth90}
{W.D. Smyth \& W.R. Peltier,}
``Three-dimensional primary instabilities of a stratified dissipative, parallel flow".
{ Geophys. Astrophys. Fluid Dyn. \/} {\bf 52}, 249--261 (1990).

\bibitem[Smyth, Klaasen \& Peltier (1988)]{Smyth88}
{W.D. Smyth, G.P. Klaasen \& W.R. Peltier,}
``Finite amplitude Holmboe waves,".
{ Geophys. Astrophys. Fluid Dyn.. \/} {\bf 43}, 181--222 (1988).

\bibitem[Smyth, \& Peltier (1991)]{Smyth91}
{W.D. Smyth \& W.R. Peltier,}
``Instability and transition in in finite amplitude Kelvin-Helmholtz and Holmboe waves".
{ J. Fluid Mech. \/} {\bf 228}, 387--415 (1991).


\bibitem[Sutherland, Caulfield \& Peltier (1994)]{Sutherland94}
{B.R. Sutherland, C.P. Caulfield \& W.R. Peltier, }
``Internal gravity generation and hydrodynamic instability".
{ J. Atmos. Sci. \/} {\bf 51}, 3261--3280 (1994).

Smyth03
\bibitem[Smyth, \& Winters (2003)]{Smyth03}
{W.D. Smyth \& K.B. Winters,}
``Turbulence and mixing in Holmboe waves".
{ J. Phys. Oceanogr.. \/} {\bf 33}, 694--711 (2003).

\bibitem[Browand \& Winant (1973)]{Browand73}
{F.K. Browand \& C.D. Winant}
``Laboratory observations of shear layer instability in a stratified fluid."
{ Boundary Layer Met. \/} {\bf 5}, 67--77 (1973).

\bibitem[Koop (1976)]{Koop76}
{C.G. Koop,} ``Instability and turbulence in a stratified shear
layer". { Tech. Rep. USCAE {\bf 134} Department of Aerospace
Engineering University of South California.} (1976)

\bibitem[Pouliquen, Chomaz \& Huerre (1994)]{Pouliquen01}
{O. Pouliquen, J.M. Chomaz \& P. Huerre,} 
``Propagating Holmboe waves at the interface between two immiscible fluids".
{ J. Fluid Mech.\/} {\bf 266}, 277--409 (1994).

\bibitem[Zhu \& Lawrence (2001)]{Zhu01}
{D.Z. Zhu \& G.A. Lawrence,}
``Holmboe's instability in exchange flows".
{ J. Fluid Mech.\/} {\bf 429}, 391--409 (2001).

\bibitem[Hogg \& Ivey (2003)]{Hogg03}
{A.M. Hogg \& G.N. Ivey,}
``The Kelvin-Helmholtz to Holmboe instability transition in stratified exchange flows".
{ J. Fluid Mech.\/} {\bf 477}, 339--362 (2003).

\bibitem[Yonemitsu, Swaters, Rajaratnam \& Lawrence (1996)]{Yonemitsu96}
{N. Yonemitsu, G.E. Swaters,N. Rajaratnam \&  G. A. Lawrence}
``Shear instabilities in arrested salt wedge flows".
{ Dyn. Atm. Oceans\/} {\bf 24}, 173--182 (1996).

\bibitem[Banks, Drazin \& Zaturska (1976)]{Banks76} 
{W.H.H. Banks, P.G. Drazin \& M.B. Zaturska}
``On the normal modes of parallel flow of inviscid stratified fluid".
{ J. Fluid Mech.\/} {\bf 75}, 149--171 (1976).

\bibitem[Howard (1961)]{Howard61} 
{ L.N. Howard,}
``A note on a paper of John Miles".
{ J. Fluid Mech.\/} {\bf 10}, 509--512 (1961).

\bibitem[Miles (1963)]{Miles63}
{J. Miles}
``On the stability of heterogeneous shear flow Part 2".
{ J. Fluid Mech.\/} {\bf 16}, 209--227 (1963).

\bibitem[Alexakis, Young \& Rosner (2002)]{Alexakis02}
{ A. Alexakis, Y. Young \& R. Rosner} ``Shear instability of
fluid interfaces: a linear analysis". { Phys. Rev. E.\/} {\bf 65},
26313 (2002).

\bibitem[Alexakis, Young \& Rosner (2004)]{Alexakis04}
{ A. Alexakis, Y. Young \& R. Rosner,} ``Weakly non-linear analysis 
of wind driven gravity waves". { J. Fluid Mech.\/} {\bf 505},
171--200 (2004).

\bibitem[Churilov (2005)]{Churilov05}
{S.M. Churilov} ``Stability analysis of stratified shear
flows with a monotonic velocity profile without inflection
points". {Izvestiya Atmospheric and Oceanic Physics\/} 
{\bf 40}, 725-736 (2004). 

\bibitem[Press(1987)]{NumericalMethods}
{W.H. Press. B.P. Flannery, S.A. Teukolosky \& W.T. Vetterling,}
{\it Numerical Recipes,}
{Cambridge University Press, (1987)}


\end{thebibliography}
\end{document}